\documentclass[journal]{IEEEtran}

\ifCLASSINFOpdf
   \usepackage[pdftex]{graphicx}
   \graphicspath{{../pdf/}{../jpeg/}}
   \DeclareGraphicsExtensions{.pdf,.jpeg,.png}
\else
   \usepackage[dvips]{graphicx}
   \graphicspath{{../eps/}}
   \DeclareGraphicsExtensions{.eps}
\fi

\date{12th April 2020}

\usepackage{subfig}
\usepackage{color}

\begin{document}

\title{Proof of Concept for Keeping Mobile Communication Channels Static With Partner Antenna With-Movements}

\author{G. Artner}

\maketitle

\begin{abstract}
\textcolor{red}{"This paper is a preprint of a paper submitted to Electronics Letters. If accepted, the copy of record will be available at the IET Digital Library".}

This work presents a proof of concept, that mobile communication channels can be kept static by the partner antenna to which the channel is formed.
The partner antenna simultaneously moves along the same trajectory with the first antenna  (\textit{with-movement}).
Experimental results are presented for movement along a straight line over a distance of several wavelengths.
Measurements were performed with 2.45\,GHz quarter-wavelength monopole antennas in an anechoic chamber.
Statistical analysis and mathematical models of the results are presented.
\end{abstract}

\section{Introduction}

Changing wireless communication channels are a serious performance inhibitor for modern wireless communication schemes.
The antenna movements in mobile and vehicular communications are a major cause of changing channels \cite{Saligheh2008,Mecklenbrauker2011}.
For example, a small scale fading environment only manifests itself as fast-fading at a receiver when the receive antenna moves through it.
Compensation of movement effects on a physical basis are already successfully used in many radio applications.
Antenna stabilizers with gyroscopic devices have been used for decades \cite{Kenyon1949}.
In \cite{Besso2007} beam aberration for transversal satellite movement is corrected.
In \cite{Jaeck2017} a projectile is equipped with an antenna arrays to keep the beam direction during rotation.
Channels can be kept static during receiver movements over distances of several wavelengths with antenna counter-movements that keep the antenna in its original position from the viewpoint of a second antenna \cite{ArtnerCSAExperiment}.
Compensation by the remote station would be beneficial for mobile devices that are typically small.

\begin{figure}[h]
\centering
\subfloat[]{\includegraphics[height=0.11\textwidth]{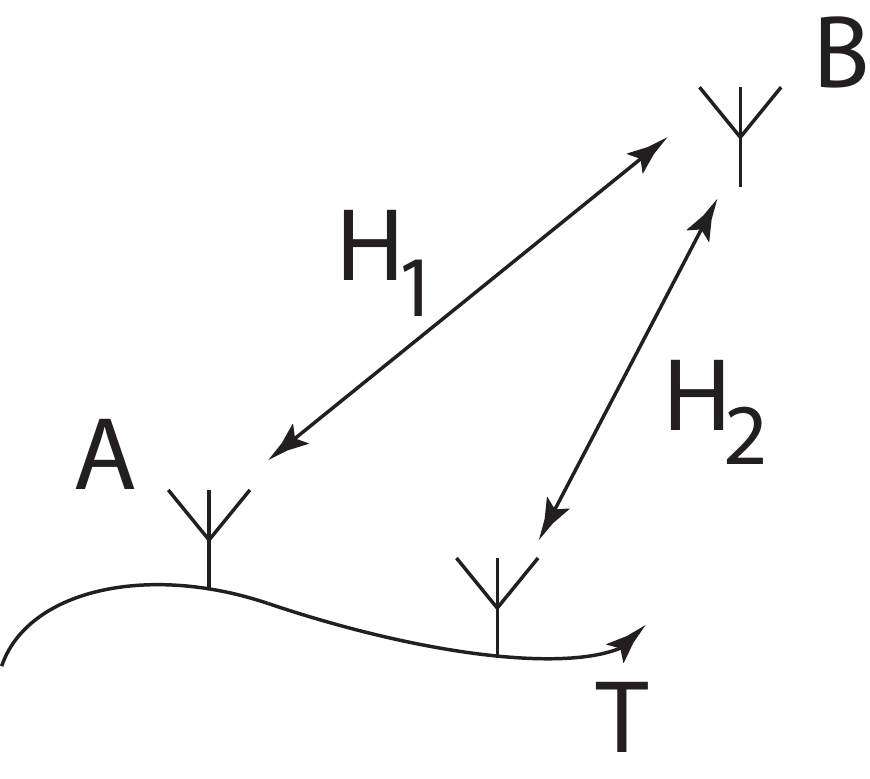}
\label{fig_volume_regular}}
\hfil
\subfloat[]{\includegraphics[height=0.11\textwidth]{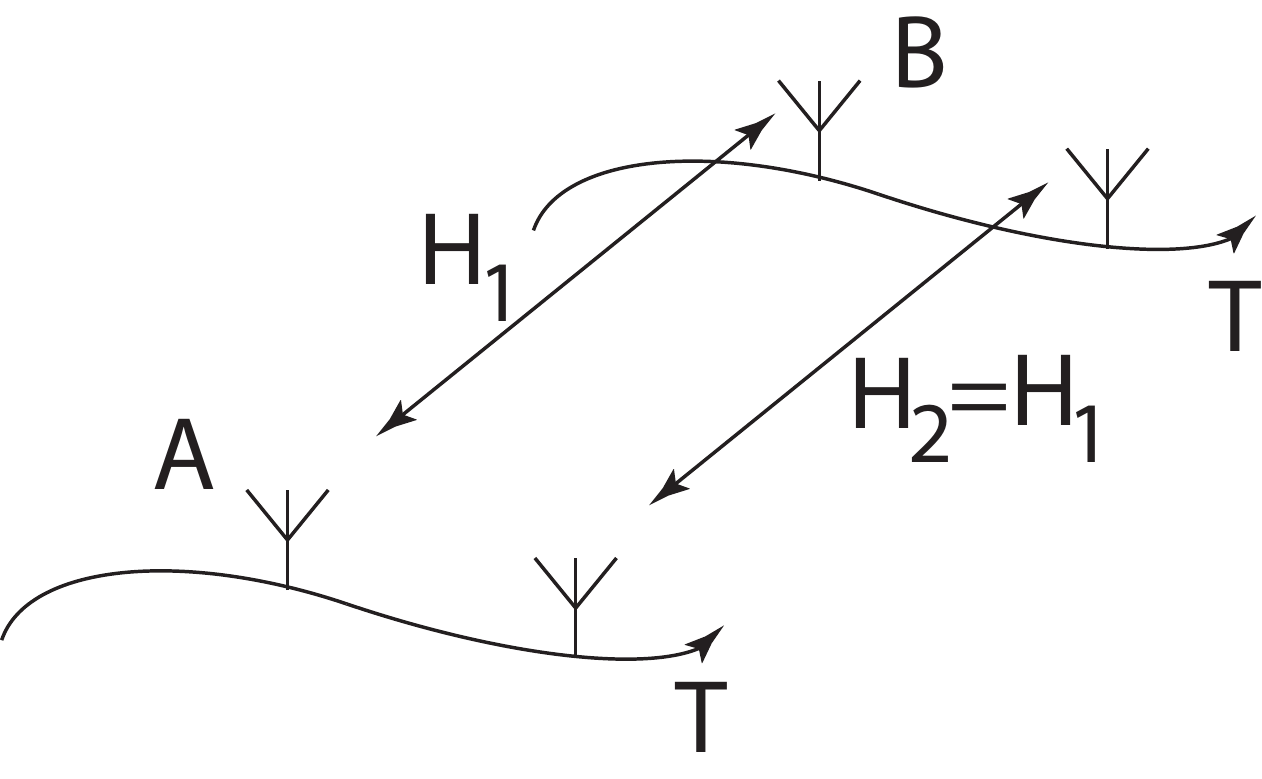}
\label{fig_volume_moved}}
\caption{Antenna A moves along a trajectory T while it communicates with antenna B through a communication channel H.\\
a) This movement changes the channel H.\\
b) B moves with A along T. This keeps the channel static as the distance between the antennas and their orientation towards each other never changes.
}
\label{fig_volume}
\end{figure}

When an antenna moves away from its position along a trajectory T, while it is connected with a second antenna through a wireless communication channel H, then the channel will in general change (see Fig.~\ref{fig_volume_regular}).
In empty space the unmoved antenna could partner with the moved antenna and simultaneously move along T to keep the orientation and distance between them.
This would subsequently keep the path-loss and phase shift of the channel static (Fig.~\ref{fig_volume_moved}).
However, to the best of the author's knowledge, such a technique that keeps mobile communication channels static on a physical basis with with-movements has not yet been investigated experimentally.
Consequently, quantitative performance assessments and channel models are not available.
The contribution of this work is to perform proof-of-concept experiments, provide a quantitative performance analysis and propose mathematical models for wireless communication channels that are kept static by with-movements of the partner antenna.

\section{Considerations}
Objects can be added to the empty space.
First, the space can be filled with homogeneous material.
Isotropic material assumptions are not required as the direction of arrival stays the same.
Second, objects that move along T do not change the channel, as long as the objects themselves or their relevant properties do not change. 
Third, objects that are translation invariant to T do not change the channel.
Radiation patterns do not need to be considered, because the antennas are always oriented towards each other at the same angle.
Note, that the previously proposed scheme in \cite{ArtnerCSAExperiment} keeps channels towards antennas at arbitrary positions static, but that this is not the case here, as only the channel between two specific antennas is kept static.
An antenna at a third position could keep its communication channel towards the first antenna static by also moving along T.

\section{Experiment}
\label{sec_measurement}

\begin{figure}[h]
\centering
\includegraphics[width=0.3\textwidth]{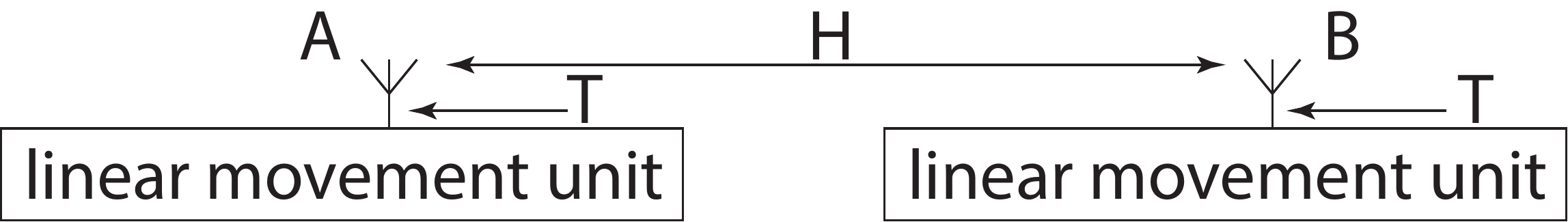}
\caption{For the experiment the two antennas are moved by CNC units. Antenna A moves along trajectory T and antenna B can simultaneously move along T to keep the channel static.}
\label{fig_isel}
\end{figure}

The feasibility of channel static partner antennas is experimentally demonstrated with the Vienna MIMO Testbed \cite{Lerch2014} in an electromagnetically shielded and absorber-lined enclosure.
For this proof-of-concept, the trajectory T is limited to linear movement.
The trajectories T are realized with two high-precision computerized numerically controlled (CNC) linear movement units with a positioning accuracy of $0.02$\,mm ($0.00016$\,$\lambda$), as is sketched in Fig.~\ref{fig_isel}.
Both antennas are quarter-wavelength wire monopole antennas for the $2.45$\,GHz industrial, scientific and medical (ISM) frequency band.
They are placed on circular aluminum ground planes with a diameter of $18$\,cm that are elevated from the movement units by $51$\,cm high posts.
The antennas are connected to a vector network analyzer (VNA) outside the chamber with coaxial cables and feedthroughs.
The VNA is trough, open, short and match (TOSM) calibrated up to the antenna ports.
The VNA measurements take some milliseconds, but antenna movement during a measurement is undesirable.
To prevent movement influences during a VNA sweep, the antennas move to new positions where they are idle for $0.2$\,s to wear off vibrations, then the VNA measurements are recorded and the antennas move to the next position and so on.
Consequently, Doppler shift \cite{Doppler1842} is not directly measured.
However, it is evident from the measured data that the residual Doppler shift is small, as the instantaneous frequency is the time derivative of the phase.
The movemetns and VNA measurements are automated with a Matlab script running on a laptop computer.

\begin{figure}[h]
\centering
\subfloat[]{\includegraphics[width=0.16\textwidth]{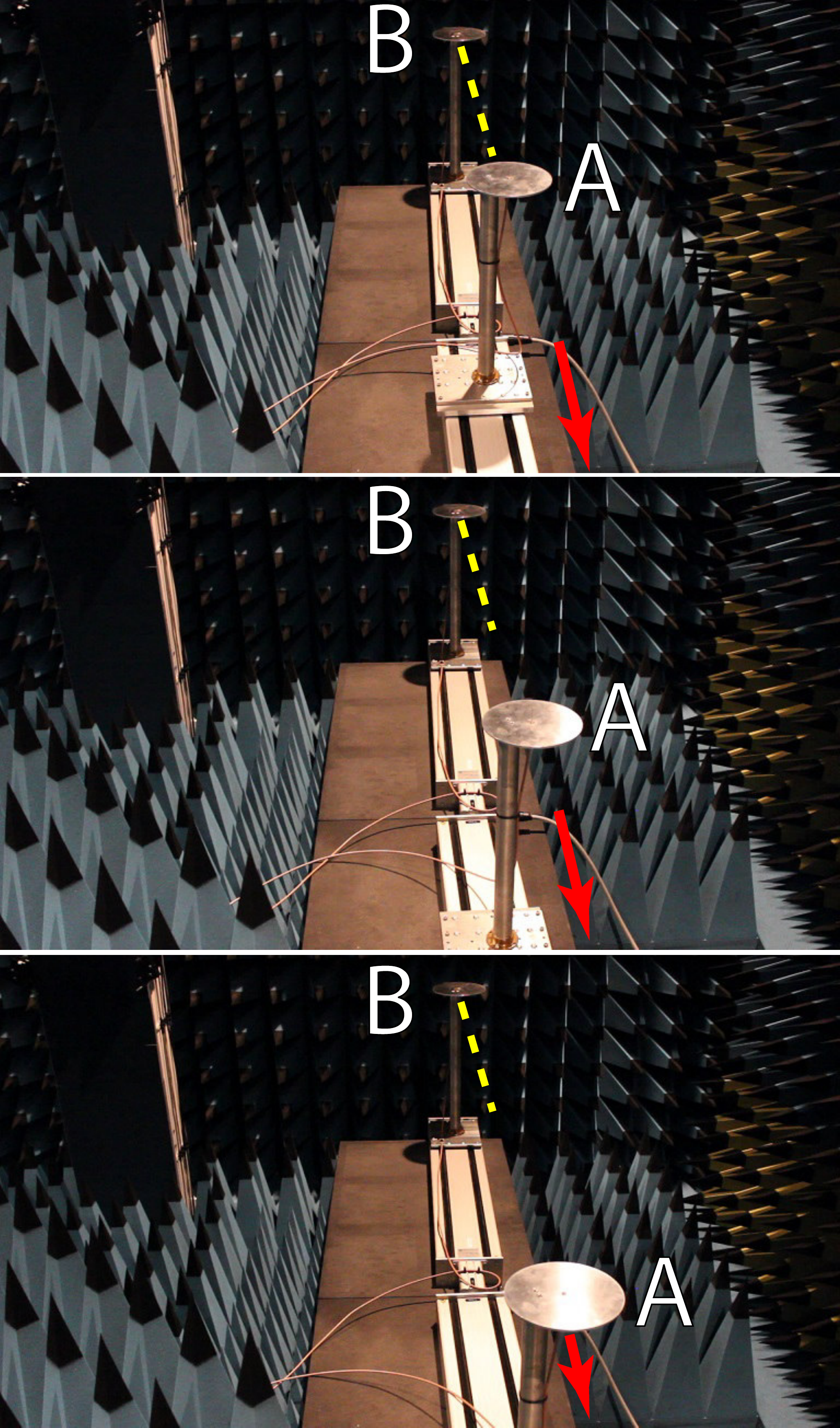}
\label{fig_move}}
~
\subfloat[]{\includegraphics[width=0.16\textwidth]{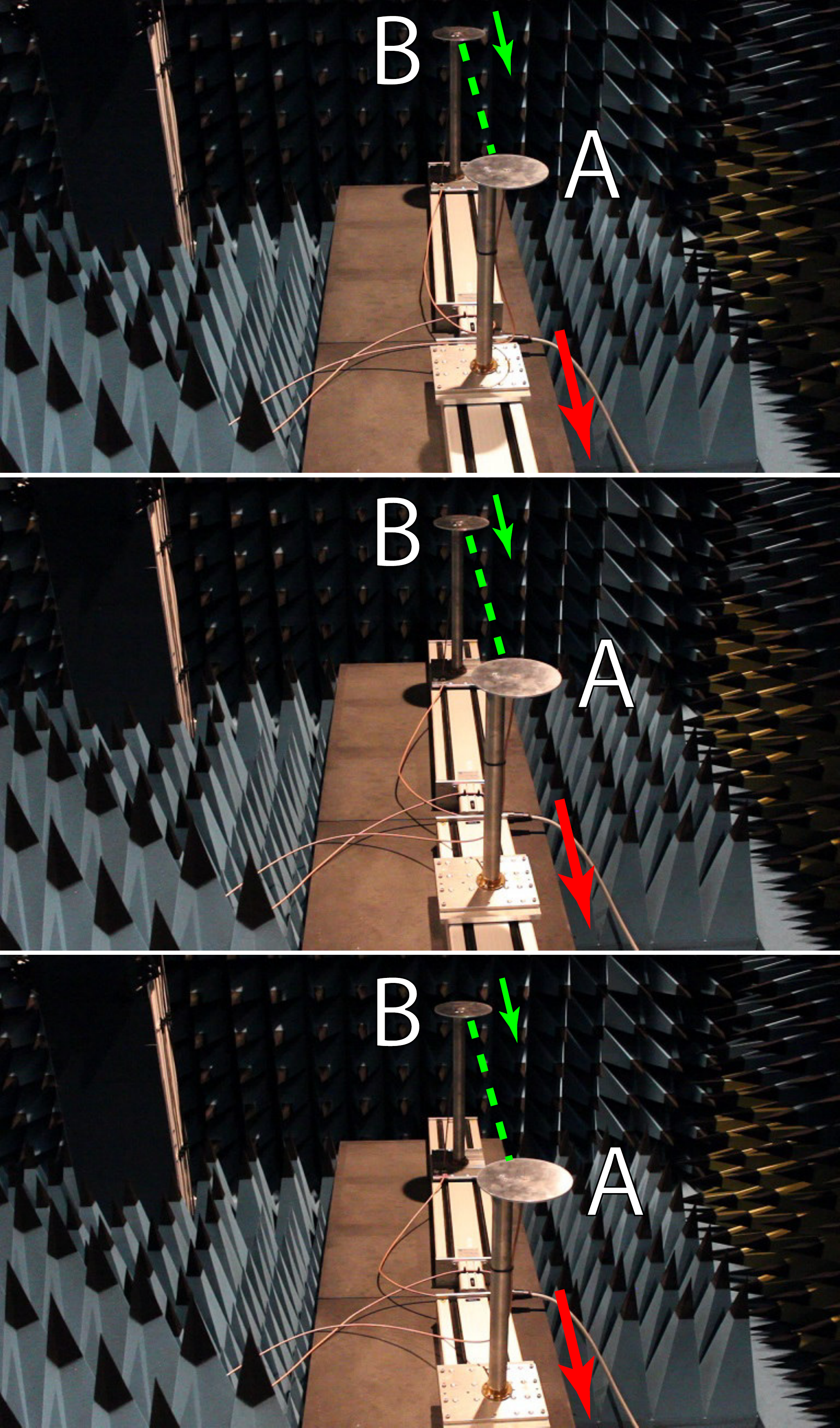}
\label{fig_compensated}}
\caption{Top to bottom: During the experiment in the anechoic chamber, antenna A moves towards the front.\\
a) The wireless communication channel between A and B changes as a result of A's movement.\\
b) Antenna B keeps the channel static by moving simultaneously with A along the same trajectory T.}
\label{fig_isel_chamber}
\end{figure}

Three measurements were performed.
First, antenna A moved away from antenna B and the movement remained uncompensated (Fig.~\ref{fig_move}).
Second, B simultaneously moved with A to keep the channel static (Fig.~\ref{fig_compensated}).
Third, both antennas remained still.
The data from the third measurement gives a reference channel without movement in the same environment and it provides an upper performance bound for compensation techniques.

\section{Results}
\begin{figure}[h]
\centering
\subfloat[]{\includegraphics[width=0.24\textwidth]{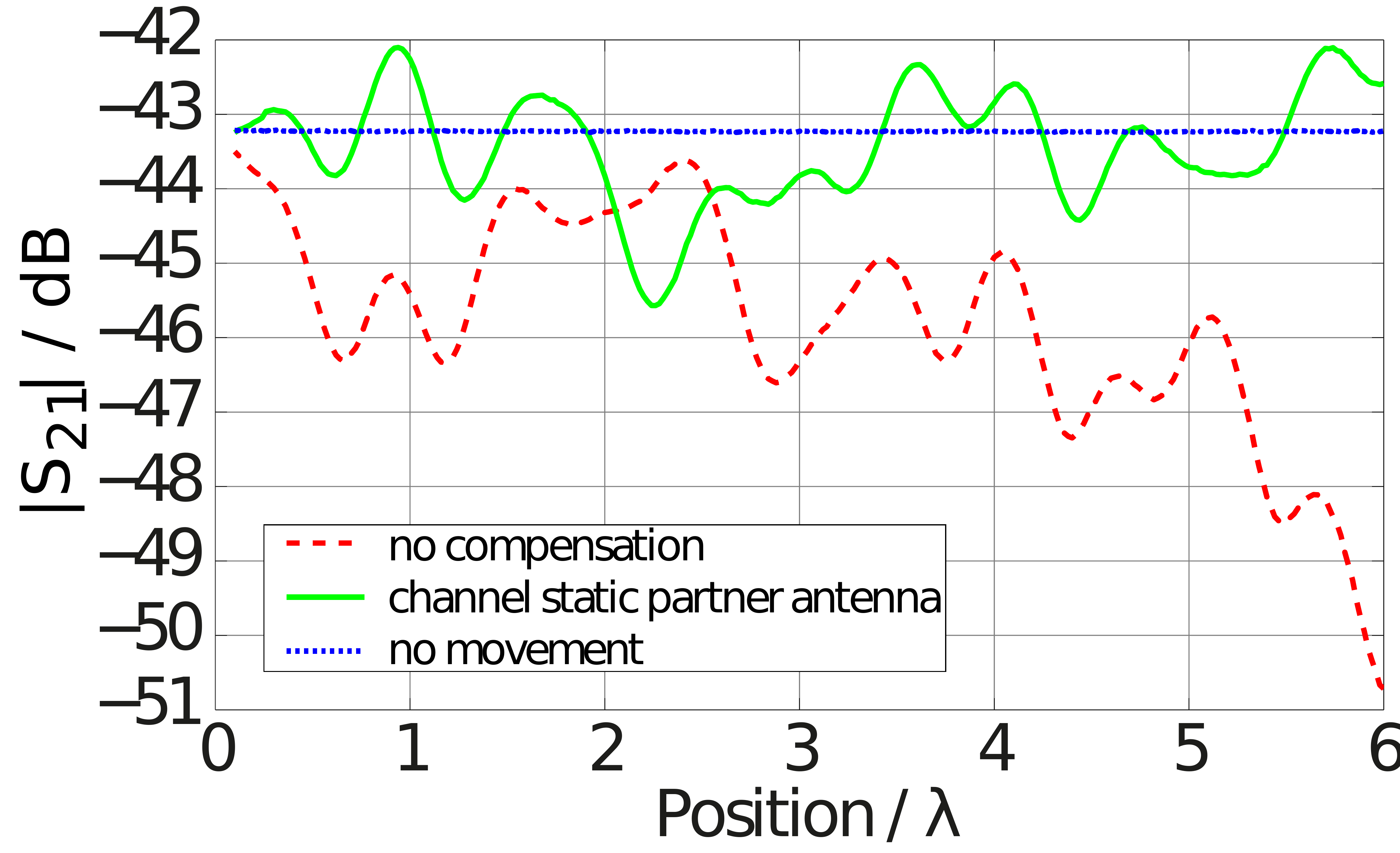}
\label{fig_results_amp}}
~
\subfloat[]{\includegraphics[width=0.24\textwidth]{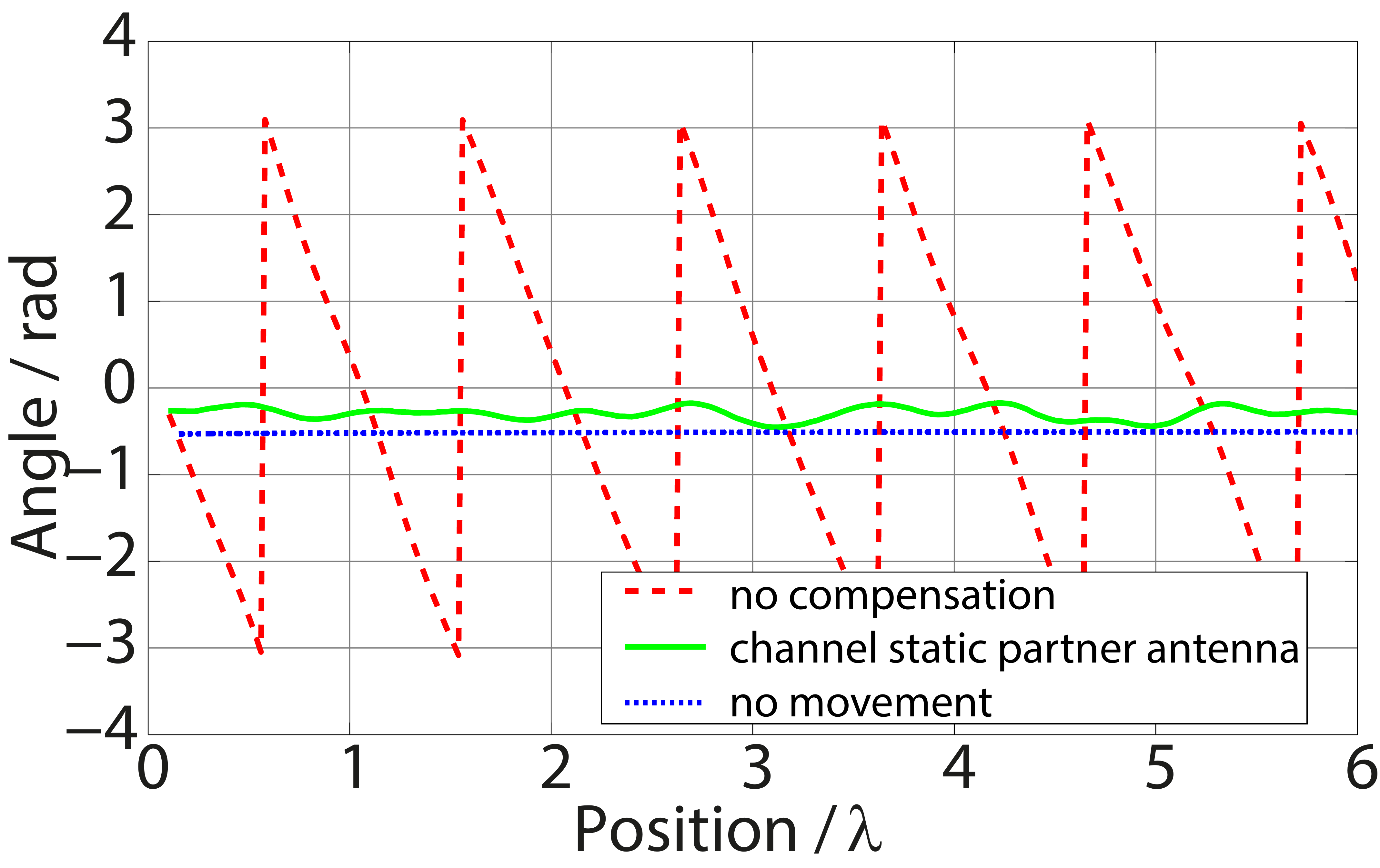}
\label{fig_results_phase}}
\caption{Measured results with the partner antenna with-movements\\
a) absolute value and \\
b) phase. The phase is wrapped at $2\pi$.}
\label{fig_results}
\end{figure}

The measured S-parameters are shown in Fig.~\ref{fig_results} and the statistical analysis is given in Tab.~\ref{tab_channel}.
The results without compensation and with the channel static partner antenna are plotted as a function of the distance that antenna A moved from its initial position in wavelengths $\lambda$.
The results without antenna movement are plotted for a similar time period as the other measurements took to provide a comparison.
The phase shift of the channel without compensation is wrapped at $2\pi$ for convenient viewing.

The anechoic environment causes only small variations of path-loss over position (Fig.~\ref{fig_results_amp}).
Without compensation, the free-space path loss increases with the distance between A and B.
The influence of antenna movement on the channel loss is quite low compared to practical environments, but it still reaches $7$\,dB without compensation.
The with-movements reduce the path loss variations to $4.5$\,dB.
The with-movements have the largest influence on the phase shift (Fig.~\ref{fig_results_phase}).
Without compensation, the phase shift increases proportionally by $2\pi$ for each $\lambda$ increase in distance between the antennas.
With with-movements the phase no longer scales with the moved distance and only varies between $0.3$\,rad ($17^\circ$) peek-to-peek and a variance of $0.0049$\,rad$^{-2}$.
Overall, the partner antenna keeps the mobile channel static within practical limits.

The remaining changes are likely the result of waves reflecting off metal parts such as the linear movement units and cables that were present during measurements and not covered by absorbers.
Inside the anechoic chamber, the channel is kept static within the same order of magnitude as with counter-movements \cite{ArtnerCSAExperiment}.
The counter-movements show a bit better performance for keeping the path-loss static and the with-movements for the phase, but performance in the anechoic chamber is close and can be considered equal for practical purposes.

\begin{table*}[t]
	\centering
		\caption{Measured mean $\mu$, maximum (peak-to-peak) and variances $\sigma^2$ of channel changes.}
		\begin{tabular}{| l | r | r | l | r | r | l|} \hline
		  & \multicolumn{3}{c|}{absolute value} & \multicolumn{3}{c |}{phase}\\ \hline
		  & $\mu$ / dB & max / dB & $\sigma^2$ / dB$^2$ & $\mu$ / rad & max / rad & $\sigma^2$ / rad$^2$\\
			\hline
			 regular (wrapped  $2 \pi$) 				& -45.8 & 7.22 & 2.1634 & -0.066  &  6.220 & 3.41\\ \hline
			 regular (not wrapped) 							& -''-  & -''- & -''-   & -18.449 & 36.172 & 109.06\\	\hline
			 channel static partner antenna 		& -43.4 & 3.47 & 0.5711 & -0.293   &  0.278 & 0.0049\\ \hline 
			 no movement 												& -43.2 & 0.04 & 4.8166 $\cdot 10^{-5}$ & -0.504   &  0.040 & $6.12 \cdot 10^{-5}$\\ \hline 

			 channel static antenna \cite{ArtnerCSAExperiment} & -43.6 & 2.45 & 0.4236   & 2.83  &  0.293 & 0.0066\\ \hline
		\end{tabular}
		\label{tab_channel}
\end{table*}

\section{Channel Model for Channel Static Partner Antennas}

Channel models are proposed for antennas that perform with-movements to keep their wireless communication channel towards moving antennas static.
The models follow those for counter-movements in \cite{ArtnerCSAExperiment}, as the results in Fig.~\ref{fig_results} and Tab.~\ref{tab_channel} show similar behavior.
The results show that the method keeps the channel static that is present at a given position or time.
As the simplest models
\begin{equation}
H(n) = H(n_0)
\label{eq_CSPA}
\end{equation}
and
\begin{equation}
H(t) = H(t_0)
\label{eq_CSPA_t}
\end{equation}
are considered, where Eqs.~\ref{eq_CSPA} and \ref{eq_CSPA_t} are the spatial and temporal formulations, respectively.
$H(n)$ and $H(t)$ are the channels at position $n$ and at time $t$, respectively, and $H(n_0)$ and $H(t_0)$ are the channels at position $n_0$ and time $t_0$, from which the channels are kept static.
The initial channels $H(n_0)$ and $H(t_0)$ are not determined within the model.
They can be obtained from other channel models or simulations, drawn from distributions or estimated from physical channels.

Based on the measurement results, the channel can be modelled as
\begin{equation}
H(n) = H(n_0) + Z(n),
\label{eq_CSPA_noise}
\end{equation}
where $H(n_0)$ is the channel at the initial position $n_0$ of the moved first antenna A at which the partner antenna B starts to perform the with-movements.
$H(n_0)$ becomes the mean of the channel, as is the case in Fig.~\ref{fig_results}.
It is kept until the second antenna no longer performs with-movements, or starts a new with-movement from a new position that results in a new initial channel $H(n_0)$.
Residual changes are modelled by the zero-mean random variable Z, which is drawn for each $n$, but keeps its variance during a static interval.
The model works for path-loss and phase, for example with $\sigma^2_{\textrm{amp}}=0.5711$\,dB$^{-2}$ and $\sigma^2_{\textrm{phase}}=0.0049$\,rad$^{-2}$ after Tab.~\ref{tab_channel}.
In case antenna B can not perform with-movements indefinitely and chooses to repeat the process with a new channel, then Eq.~\ref{eq_CSPA_noise} results in a channel model that is stationary within an interval, analogous to \cite{ArtnerCSAExperiment}.
However, models for different scenarios and environments have to be determined in future work.

\section{Conclusion}

This work demonstrates that mobile communication channels can be kept static on a physical basis by a with-movement at the remote antenna.
Small devices like smart phones, where the antenna might have difficulty to perform a counter-movement \cite{ArtnerCSAExperiment} due to size limitations, can offload the movement to a larger partner antenna such as a base station.
The free-space path loss and the phase shift no longer increase proportionally with the distance between the antennas and is instead decreased to residual changes with a variance of $0.57$\,dB$^{-2}$ and $0.0049$\,rad$^{-2}$, respectively.
The measured performance is similar to channel static antennas that perform counter-movements \cite{ArtnerCSAExperiment}.
This work motivates further research.
The performance in different scenarios and environments needs to be evaluated.
The partner antenna needs two key features: A way to obtain T and a mechanism to move along T.
Both problems require inventive solutions.

\section{Conflicts of Interest}
Technische Universit\"at Wien, the author's employer at the time of invention, has filed a patent \cite{ArtnerPCSA}.

{The author thanks R.~Langwieser and S.~Pratschner, both of Technische Universit\"at Wien, Vienna, Austria, for their help with the experimental work.
He thanks C.F.\,Mecklenbr\"auker and colleagues from COST CA15104 IRACON for valuable discussions.}
\vskip5pt
\noindent G. Artner (\textit{T\"UV AUSTRIA Services GmbH, Vienna, Austria;} formerly \textit{Institute of Telecommunications, Technische Universit\"at Wien, Vienna, Austria})
\vskip3pt
\noindent E-mail: gerald.artner@nt.tuwien.ac.at

\end{document}